\documentclass[sigconf]{acmart}

\AtBeginDocument{%
  }

\usepackage{array}
\usepackage{xcolor} 
\usepackage{booktabs} 
\usepackage{colortbl}

\newcommand{\edit}[1]{\textcolor{black}{#1}}

\copyrightyear{2026}
\acmYear{2026}
\setcopyright{cc}
\setcctype{by-nc-nd}
\acmConference[CHI EA '26]{Extended Abstracts of the 2026 CHI Conference on Human Factors in Computing Systems}{April 13--17, 2026}{Barcelona, Spain}
\acmBooktitle{Extended Abstracts of the 2026 CHI Conference on Human Factors in Computing Systems (CHI EA '26), April 13--17, 2026, Barcelona, Spain}
\acmDOI{10.1145/3772363.3798545}
\acmISBN{979-8-4007-2281-3/2026/04}

\begin{document}

\title{Balancing Openness and Safety: Central and Peripheral Governance Practices in the Lesbian Subreddit Ecosystem}

\author{Yan Xia}
\affiliation{%
  \institution{Clemson University}
  \city{Clemson}
  \country{USA}}
\email{xia5@clemson.edu}

\author{Sushmita Khan}
\affiliation{%
  \institution{Clemson University}
  \city{Clemson}
  \country{USA}}
\email{sushmik@clemson.edu}

\author{Naiyah Lewis}
\affiliation{%
  \institution{Clemson University}
  \city{Clemson}
  \country{USA}}
\email{naiyahl@clemson.edu}

\author{Jinkyung Katie Park}
\affiliation{%
  \institution{Clemson University}
  \city{Clemson}
  \country{USA}}
\email{jinkyup@clemson.edu}

\renewcommand{\shortauthors}{Xia et al.}

\begin{abstract}
Online LGBTQ+ communities face a persistent tension: remaining visible to welcome newcomers while protecting members from harassment. This challenge is particularly acute for lesbian communities on Reddit, which operate not as isolated groups but as an interconnected ecosystem. We examine how this tension is negotiated across the lesbian subreddit ecosystem ($N=29$) by combining network analysis of cross-subreddit links with a qualitative thematic analysis of 167 subreddit rules. Our findings show a functional division of governance labor between central (34\%) and peripheral subreddits (66\%). While all communities share a baseline of safety regulations, central subreddits prioritize content curation and feed quality to support a large, public-facing audience, whereas peripheral subreddits emphasize boundary maintenance and participation control to protect smaller, identity-specific niches. These findings challenge monolithic moderation approaches and highlight the need for ecosystem-aware design. We argue that effective moderation requires role- and context-sensitive tools supporting visibility and safety across interconnected spaces.

\end{abstract}

\ccsdesc[500]{Human-centered computing~Collaborative and social computing}
\ccsdesc[500]{Human-centered computing~Human computer interaction (HCI)}

\keywords{Online communities, Platform Governance, Lesbian Online Communities, Reddit, Network analysis, Qualitative Thematic Analysis}

\maketitle

\section{Introduction}
\begin{figure*}[ht]
    \centering
    \includegraphics[width=\textwidth]{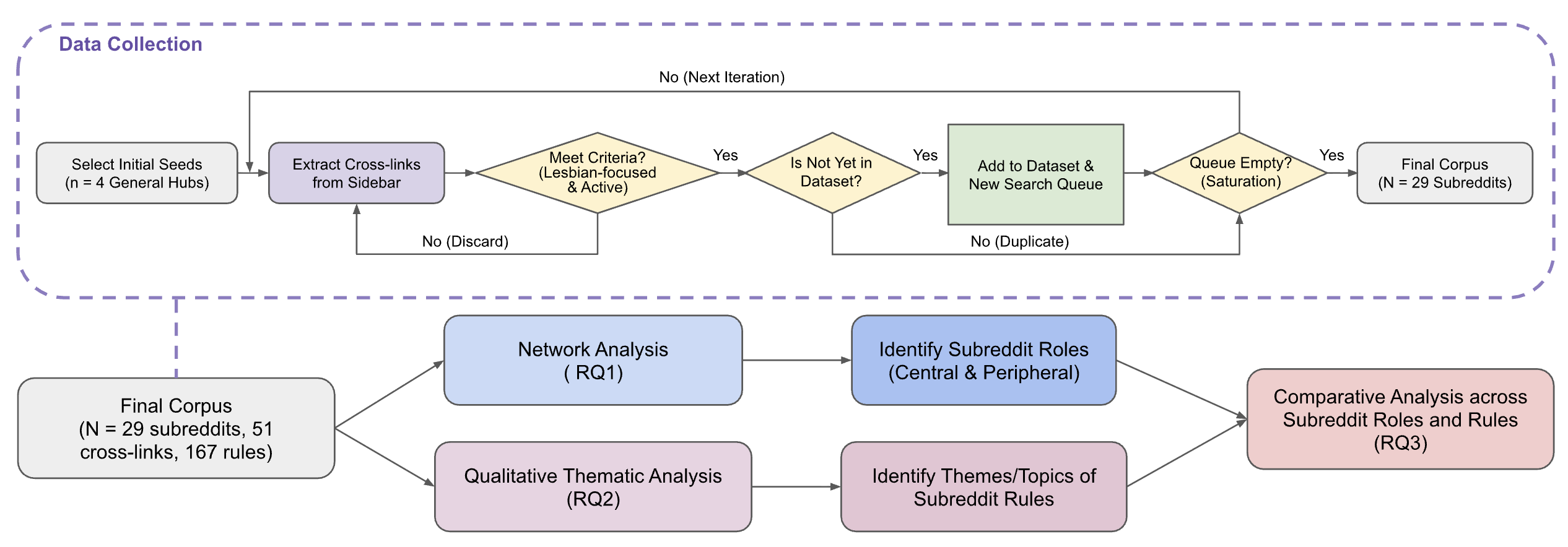}
    \Description{The upper part illustrates the iterative snowball sampling procedure. Beginning with four seed subreddits, we collected cross-links to other subreddits and filtered candidates for relevance. This process was repeated until theoretical saturation, resulting in a final corpus of 29 communities. The lower part outlines our method design. We conducted parallel network and qualitative content analyses to identify structural roles and governance practices features, respectively. These findings were synthesized to reveal how governance strategies diverge between central and peripheral subreddits.}
    \caption{The pipeline of our methods for data collection and data analysis. We combined network analysis with qualitative thematic analysis of subreddit rules to map distinct governance practices to specific ecosystem roles.}
    \label{fig:methods}
\end{figure*}



For LGBTQ+(lesbian, gay, bisexual, transgender, queer, and having other sexual orientations and gender identities) individuals~\cite{blondeel-2016, zeeman-2018}, online communities are not merely discussion forums; they serve as vital infrastructure for identity work, social connection, and survival~\cite{teblunthuis-2022, craig-2021}. Many users even feel safer and more supportive in these spaces than their offline counterparts, particularly when navigating stigma or isolation~\cite{mcinroy-2019}. Yet, this safety and support is precarious, especially for lesbians, who face persistent harassment and harm online~\cite{friedman2007lesbians, vashistha2019}. Simultaneously, lesbian communities are increasingly facing tension between cohesive identity politics and more fluid queer frameworks~\cite{soares-2024}, complicating how boundaries, participation, and safety are negotiated. 

As a result, these digital spaces must balance a fundamental desgin tension: remaining open enough to welcome newcomers while being closed enough to protect vulnerable users from harassment, marginalization, and violence~\cite{haimson-2025}. Prior work shows that no single community can meet all user needs~\cite{teblunthuis-2022}. Instead, communities differentiate and specialize, forming ecosystems that distribute social and informational labor across interconnected spaces~\cite{zhu2014, teblunthuis-2025}. In return, users navigate across multiple distinct groups to balance their need for belonging with distinctiveness~\cite{zhang-2024}.

Despite this ecosystem reality, research on online community governance has largely focused on moderation and policy at the level of individual communities~\cite{kiene2016, jhaver-2019}.  Governance, which is enacted through explicit rules, moderation enforcement, and normative boundaries, plays a central role in defining community scope and protecting niche identities~\cite{fiesler-2018, kiene2019volunteer, leibmann-2025}. While prior work has mapped Reddit’s governance landscape and examined adaptation to competition and growth~\cite{fiesler-2018, teblunthuis-2025}, we still lack an understanding of how a community’s position within an ecosystem influences its governance strategies. In particular, it remains unclear whether and how central (hub-like) and peripheral (niche) communities differ in balancing openness, safety, and identity protection.

To address this gap, we analyzed the governance practices of lesbian-focused subreddits.  We combined network analysis of cross-links~\cite{kumar-2018} (see screenshots in Appendix, subreddits mention others on sidebars) with a qualitative thematic analysis of community rules to examine how governance practices align with a subreddit's ecosystem role. This approach offers a structural lens to understand how marginalized communities collectively manage identity, participation, and safety within an interconnected ecosystem. 
This work is guided by the following research questions:
\begin{itemize}
    \item \textbf{RQ1}: How are central and peripheral subreddits structurally positioned within the lesbian subreddit ecosystem?
    
    \item \textbf{RQ2}: What governance practices are articulated through lesbian subreddit rules across the ecosystem?

    \item \textbf{RQ3}: How do governance practices differ between the central and peripheral lesbian subreddits?
    
\end{itemize}

By examining governance practices at the ecosystem level, this work contributes to CHI on online safety, community governance, and marginalized users by demonstrating that effective protection and inclusion emerge from interdependent governance strategies across a network of communities.

\section{Methods}

\paragraph{2.1 Data Collection}
To capture the network of mutual references among lesbian-specific subreddits, we employed a purposive sampling strategy and iterative snowball sampling~\cite{campbell-2020, goodman-1961} (See Figue \ref{fig:methods}). 

We constructed the lesbian subreddit ecosystem using an iterative snowball sampling approach. We began with four large ($>$20k members), general-interest (i.e., broad discussion-focused), and non-NSFW (not safe for work) subreddits (i.e., \texttt{r/actuallesbians}, \texttt{r/LesbianActually}, \texttt{r/ActuallyLesbian}, and \texttt{r/lesbiangang}) and followed moderator-curated sidebar cross-links to identify related communities (See Figures in Appendix). Sidebar links were used as the primary connection metric because they represent intentional endorsements between communities~\cite{hessel2016science, kumar-2018}. We included only subreddits explicitly focused on lesbian/WLW (Women Loving Women) identities and excluded banned or unmoderated communities. Sampling continued until no new valid subreddits were identified. This workflow (see Figure \ref{fig:methods}) yielded a corpus of 29 lesbian-focused subreddits, including 167 subreddit rules and 51 cross-community links for analysis.

\paragraph{2.2 Data Analysis}

To understand the network shape and governance practices of the lesbian subreddit ecosystem, we conducted a network analysis of cross-community links and a qualitative thematic analysis of the subreddit rules. \edit{This approach allowed us to capture both the shape of the ecosystem and the regulatory practices enacted within individual communities.}

\paragraph{2.2.1 Network Analysis.}

To understand the structural topology of this ecosystem, we constructed a directed network of the 29 subreddits, utilizing the 51 cross-links found in sidebar widgets as edges. We analyzed this structure to determine how different communities interact, focusing on two main factors. We measured degree centrality to identify which communities act as popular destinations (in-degree) and which act as signposts directing users to other spaces (out-degree).

Center-periphery structure~\cite{teblunthuis-2025}: To distinguish central subreddits from outliers, we analyzed the connectivity components. We defined ``Central'' communities as those belonging to the ecosystem's connected component (having at least one tie). In contrast, we defined ``Peripheral'' subreddits as strictly isolated nodes with no sidebar links to the rest of the group. Visualizing this network (see Figure \ref{fig:radar_plots} left) using the Fruchterman-Reingold algorithm~\cite{fruchterman-1991} reveals a highly fragmented landscape.

\edit{Although centrality is often treated as a continuous property, we operationalize a binary distinction between central and peripheral communities to capture qualitatively different ecological roles. Prior work on online community ecosystems shows that communities persist by specializing to avoid competition and by occupying complementary roles, rather than by positioning themselves along a single popularity dimension~\cite{teblunthuis-2025}. From this perspective, the key distinction is whether a community is structurally embedded in inter-community ties that shape its governance and participation dynamics, or operates in relative isolation~\cite{granovetter-1985,uzzi-1997,lazega-1995,dyer-1998,teblunthuis-2025}. On Reddit, moderator-curated sidebar cross-links function as formal signals of relatedness and navigational pathways between communities. We therefore treat participation in the connected component as indicating ecosystem-level embeddedness, and isolation as reflecting a distinct niche role, justifying a categorical distinction between central and peripheral communities as an analytic lens.}

\paragraph{2.2.2 Qualitative Thematic  Analysis.}

To analyze governance practices, we conducted an inductive qualitative thematic analysis~\cite{hsieh-2005} of 167 subreddit rules. Following established inductive qualitative analysis procedures~\cite{fiesler-2018, strauss-1998}, 
a primary coder engaged in iterative open coding to summarize each rule into descriptive tags, which were then grouped and refined to eliminate redundancy and reach thematic saturation. These categories were organized into a two-level analytical framework, with themes capturing broad governance goals (e.g., safety and risk mitigation) and topics representing specific moderation strategies. To ensure analytic validity, a second coder reviewed the framework and coding application, and disagreements were resolved through discussion. The Principal Investigator finalized the codebook, resulting in a taxonomy of five themes and eleven topics, which we used to compare governance practices between central and peripheral subreddits.
\section{Findings}
\begin{figure*}[ht]
    \centering
    \includegraphics[width=\textwidth]{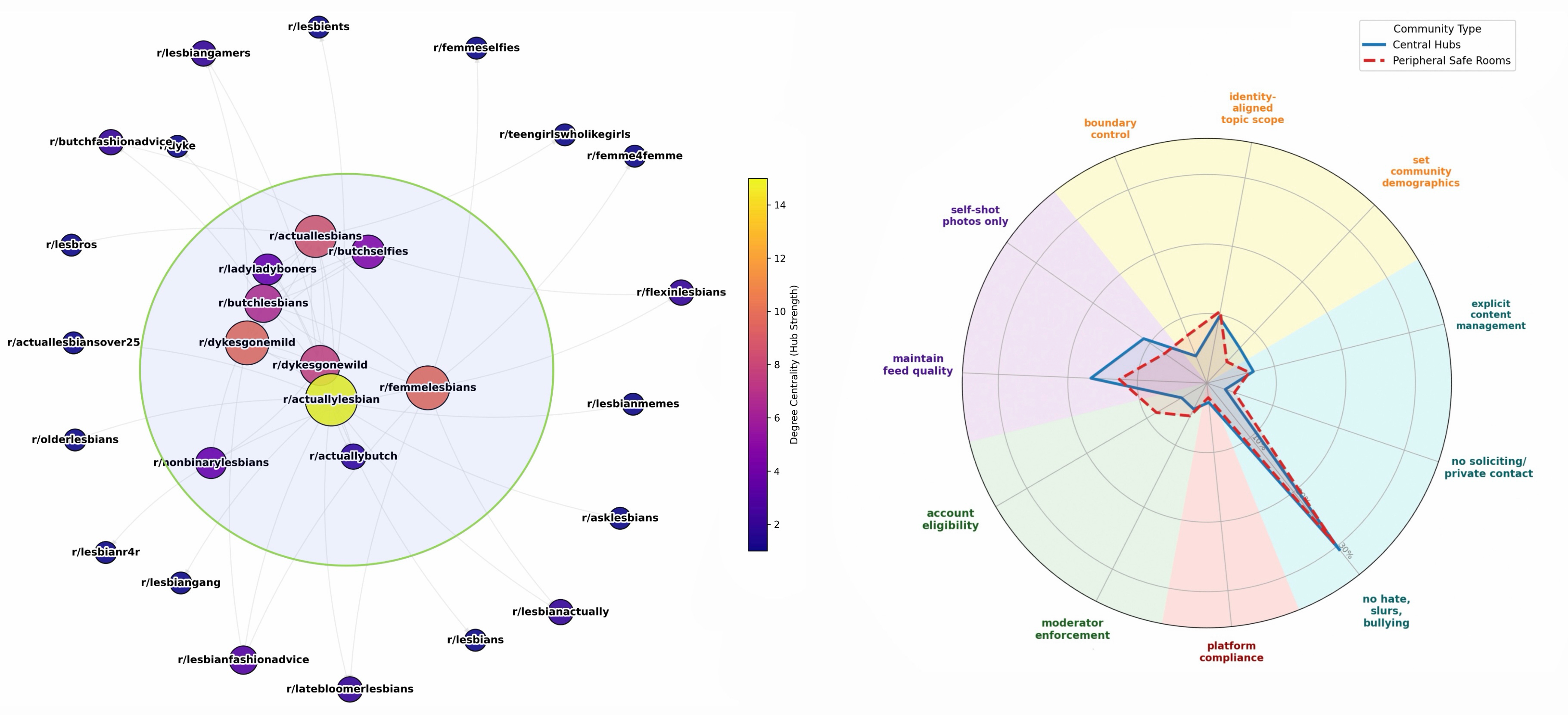}
    \Description{The left chart displays the subreddit ecosystem network generated by cross-links using the Fruchterman-Reingold algorithm.  The right chart displays Topics. The charts use colored background sectors to indicate rule categories.}
    \caption{Left: Subreddit Ecosystem Network generated by cross-links. The subreddits within the green circle are central, outliers are peripheral. Right: Topics detailed rules show how central and peripheral subreddits focus. Blue, yellow, purple, pink, green sectors correspond to rule themes \textit{Safety and Risk Mitigation, Identity and Boundary Management, Content Creation and Quality Control, General Instruction, Participation Control and Reinforcement,} respectively.}
    \label{fig:radar_plots}
\end{figure*}

\paragraph{3.1 Centrals vs. Peripherals}
Network analysis reveals that while 34\% ($n=10$) of the ecosystem functions as central hubs, the majority (66\%) ($n=19$) of subreddits are peripheral islands (see Figure \ref{fig:radar_plots} left). Notably, the subreddits occupying the central positions are mostly general-purpose. These subreddits act as the backbone of the ecosystem, serving as entry points and central hubs where users can engage in broad discussions. The subreddits on the periphery are more focused on age-specified groups or particular hobbies. These subreddits cater to specific demographic groups, such as teenagers and older lesbians, or emphasize shared interests, their low connectivity suggests a more insular approach, possibly designed to maintain tight-knit, interest-specific communities.

\paragraph{3.2 Rules across the Subreddit Ecosystem}
Our inductive qualitative thematic analysis of 167 rules across the lesbian subreddit ecosystem reveals the big goals and specific methods used by moderators to manage these digital spaces. This system of rules is structured into five themes representing broad goals and 11 topics detailing the specific methods for governance practices (see Table \ref{tab:governance_taxonomy}). 

\definecolor{GenInstBG}{HTML}{FFE2E0}
\definecolor{SafeRiskBG}{HTML}{DFF7F8}
\definecolor{IdentityBG}{HTML}{FDFCD6}
\definecolor{ContentBG}{HTML}{F3E4F5}
\definecolor{ParticBG}{HTML}{E8F5E9}

\begin{table}[ht]
\centering
\footnotesize
\renewcommand{\arraystretch}{1.3}
\setlength{\tabcolsep}{1.5pt}
\caption{The Themes and Topics of Lesbian Subreddit Rules.}
\label{tab:governance_taxonomy}
\begin{tabular}{@{}
    >{\raggedright\arraybackslash}p{0.18\linewidth}
    >{\raggedright\arraybackslash}p{0.32\linewidth}
    >{\raggedright\arraybackslash}p{0.46\linewidth}
@{}}
\toprule
\textbf{Themes} & \textbf{Topics} & \textbf{Topic Descriptions} \\
\midrule
{\textbf{\cellcolor{SafeRiskBG}Safety and Risk Mitigation}} &
\cellcolor{SafeRiskBG}{\textbf{No hate, slurs, bullying, or discrimination}} &
\cellcolor{SafeRiskBG}Strictly prohibits transphobia, racism, trolling, and personal attacks, enforcing a ``zero tolerance policy for hate'' and requiring users to be civil. \\
\cellcolor{SafeRiskBG}& \cellcolor{SafeRiskBG}{\textbf{No soliciting or further private contact}} &
\cellcolor{SafeRiskBG}Bans ``dating/chat/hookup'' posts, asking for private messages, sharing contact details in public, and promoting Discord servers. \\
\cellcolor{SafeRiskBG}& \cellcolor{SafeRiskBG}{\textbf{Explicit content management}} &
\cellcolor{SafeRiskBG}Governs the posting and tagging of explicit content. \\
\addlinespace[2pt]
\cellcolor{IdentityBG}{\textbf{Identity and Boundary Management}} &
\cellcolor{IdentityBG}{\textbf{Set community demographics}} &
\cellcolor{IdentityBG}Sets specific demographics for the community based on identity and common interest. \\
\cellcolor{IdentityBG}& \cellcolor{IdentityBG}{\textbf{Identity-aligned topic scope}} &
\cellcolor{IdentityBG}Specifies the topics that promote the identity of the community and prohibit unrelated topics. \\
\cellcolor{IdentityBG}& \cellcolor{IdentityBG}{\textbf{Boundary control}} &
\cellcolor{IdentityBG}Explicitly bans men, straight couples to keep the focus on the community's intent. \\
\addlinespace[2pt]
\cellcolor{ContentBG}{\textbf{Content Creation and Quality Control}} &\cellcolor{ContentBG}
{\textbf{Self-shot photos only: no AI or stolen content}} &
\cellcolor{ContentBG}Post original, self-shot photos in designated threads/times only. No AI or stolen content allowed. \\
\cellcolor{ContentBG}& \cellcolor{ContentBG}{\textbf{Maintain feed quality \& searchability}} &
\cellcolor{ContentBG}Excludes excessive promotional content, repetitive submissions, reposts, crossposts, and censored titles that disrupt search functionality. \\
\addlinespace[2pt]
\cellcolor{GenInstBG}{\textbf{General Instructions}} &
\cellcolor{GenInstBG}{\textbf{Platform compliance}} &
\cellcolor{GenInstBG}Adhere to all Reddit site-wide rules, Terms of Service, and subreddit rules and FAQs. \\
\addlinespace[2pt]
\cellcolor{ParticBG}{\textbf{Participation Control and Reinforcement}} &
\cellcolor{ParticBG}{\textbf{Account eligibility \& verification}} &
\cellcolor{ParticBG}Details the prerequisites for an account to post, including age limits, karma thresholds, and the specific verification process. \\
\cellcolor{ParticBG}& \cellcolor{ParticBG}{\textbf{Moderator enforcement \& discretion}} &
\cellcolor{ParticBG}Outlines the consequences of breaking rules (3 strikes, bans) and the authority moderators have to remove accounts and content. \\
\bottomrule
\end{tabular}
\end{table}

Our analysis shows that this ecosystem is not simply a list of rules, but rather a multi-layered protection system tailored to vulnerable groups. The ``Safety and Risk Mitigation'' theme reveals a layered defense system designed to mitigate varying degrees of harm. Topics like ``No Hate, Slurs, or Bullying'' belong to this theme because, in this ecosystem, hate speech (e.g., transphobia, racism) is treated as a threat to safety rather than a mere violation of politeness. Rules under the ``No Soliciting or Further Private Contact'' topic aim to mitigate the risk of off-platform harm by banning hookup requests and soliciting direct messages, protecting users from potential predatory behavior in one-on-one interactions. The rules under the``Explicit Content Management'' theme protects users from non-consensual exposure to sexually explicit imagery or protects the whole subreddit from unwanted labels. The ``Identity and Boundary Management'' theme aim to operate on two distinct levels to preserve the community's integrity. First, at the user level, rules regarding the ``Boundary Control'' and ``Set Community Demographics'' topics explicitly codify membership, such as banning men or straight couples to prevent the polarization of the lesbian-centric experience. Secondly, the rules of the ``Identity-Aligned Topic Scope'' topic enforce a specific topical focus, maintaining a cohesive cultural space.

\paragraph{3.3 The Difference of Subreddit Rules between Center and Periphery}
Our comparative analysis reveals that themes and topics in lesbian subreddit rules differ between central subreddits and peripheral subreddits (see Figure \ref{fig:radar_plots} right chart). \edit{While structural connectivity exists on a continuum, this distinction highlights a meaningful break between communities that participate in shared navigational and mutualistic dynamics and those that function as specialized niche spaces.}

Central subreddits focus more on the ``Content Creation and Quality Control'' theme(see Figure \ref{fig:radar_plots} right chart purple sector) than peripheral subreddits. At the topic level, this manifests as a high frequency of the ``Maintain Feed Quality'' and ``Self-Shot Photos Only'' topic rules. This suggests that central subreddits focus more on keeping the content feed clean and easy to read for a large audience. Notably, while both groups regulate identity, they regulate it differently. Central subreddits focus more on the ``Set Community Demographics'' topic (defining who we are generally) more than peripheral groups, but have fewer rules regarding explicit ``Boundary Control'' topic than peripheral subreddits. This indicates a strategy of ``positive definition''~\cite{mummendey-2000}, broadly signaling the community's identity to welcome newcomers without imposing strict exclusionary gates a lot. In contrast, peripheral subreddits focus more on the ``Participation Control and Reinforcement'' theme (see Figure \ref{fig:radar_plots} right chart green sector). These communities rely more on gating methods, such as the rules of ``Account Eligibility'' (e.g., karma thresholds, verification) and strict ``Moderator Enforcement'' topics. The peripheral subreddits place a higher focus on the ``Boundary Control'' topic (defining who is not allowed, e.g., explicitly banning men or straight couples) while focusing less on general demographic statements. This reflects an ``ecological niche'' strategy: because they are smaller and more specific, peripheral communities must actively police their borders to prevent context collapse, focus more on the integrity of the user base over the flow of content.

\section{Discussions}
Our findings reveal 
a functional division of moderation where central and peripheral subreddits adopt distinct regulatory strategies to balance the openness and safety~\cite{haimson-2025}, by specializing in either content curation or identity protection.

\paragraph{4.1 The Ecosystemic Level Resolution}
Prior work often frames the tension between ``openness'' and ``safety'' as a conflict that individual communities must resolve internally~\cite{haimson-2025}. \edit{Our findings align with ecological accounts of online communities that emphasize specialization and mutualism over uniform scaling. As prior work argues, online community ecosystems stabilize when communities occupy complementary roles rather than competing along the same dimensions~\cite{teblunthuis-2022}.} We show that lesbian subreddits differentiate and specialize into specific niches~\cite{zhu2014, teblunthuis-2025} that collectively balance visibility, safety, and social support. 
Central subreddits function as high-visibility hubs that support broad participation and information exchange, but face challenges of scale, including noise and limitations on deeper peer socialization. In contrast, peripheral communities leverage their relative isolation to provide the homophily, intimacy, and specialized identities that larger hubs cannot sustain simultaneously~\cite{teblunthuis-2025}. Together, these differenceiated roles form an ecosystem that enables users to move cross spaces as their needs evolve, supporting identity development through participation in multiple and interconnected communities~\cite{haimson-2018}. These findings suggest that online communities should be understood not as isolated units, but as interdependent components of a broader governance ecology.

\paragraph{4.2 Identity as Infrastructure}
A critical finding of this work is the identification of Identity and Boundary Management as a distinct strategy for governance. In general-interest online communities, moderation typically focuses on behavior (e.g., civility, spam). In the lesbian ecosystem, moderation focuses heavily on identity (e.g., set demographic expectations). The critical feature of an ethnic or identity-based group is the social boundary that defines the group rather than the ``cultural stuff'' it encloses~\cite{freedman-1970}. Because the term ``lesbian'' is currently a ``term in dispute'' caught between cohesive identity politics and fluid queer frameworks~\cite{soares-2024}, subreddits act as boundary markers to prevent context collapse~\cite{costa-2018}. Central subreddits tend to utilize positive definition to signal a broad identity and welcome a diverse range of users, whereas peripheral subreddits employ strict boundary control regulation to maintain a cohesive cultural space for their specific niche~\cite{nakamura-2002}. This allows users to navigate across multiple communities to balance their need for belonging with their need for distinctiveness~\cite{zhang-2024}, assembling safer, context-sensitive identity expressions across their personal social media ecosystems~\cite{devito-2018, marwick-2010}.


\paragraph{4.3 Safety and Risk Mitigation in Depth}
In marginalized digital spaces, safety rules are not merely about politeness; they function as survival infrastructure shaped by asymmetric power relations~\cite{scheuerman-2018}. Power is embedded within the architecture of digital spaces~\cite{bertol-1996}, thus influencing how marginalized groups interact within these spaces. Our analysis shows that prohibitions against hate, slurs, and harassment are near-universal, reflecting persistent threats from both external hostility and internal conflict. In addition, the lesbian Reddit ecosystem also regulates participation control and reinforcement aspects to manage the balance between openness and safety~\cite{haimson-2025}. Rules governing account eligibility, verification, and moderator discretion determine who is permitted to participate and under what conditions, thereby shaping the boundaries of queer space. These rules collectively allow communities to regulate access in response to harassment, infiltration, or exploitation, while maintaining openness at the ecosystem level. As such, safety in this ecosystem emerges from layered and role-specific governance strategies distributed across communities. 

\paragraph{4.4 Implications for Design}
\edit{Prior work on lesbian-oriented and queer women’s platforms highlights that such spaces are rarely designed as monolithic communities, but instead support diverse practices of identity exploration, relationship-building, and privacy management across different platform contexts. For example, studies of sexual-minority women’s use of online dating platforms in China show that users strategically manage visibility and self-disclosure to balance connection-seeking with safety concerns, often relying on implicit cues and platform affordances to avoid unwanted exposure~\cite{cui-2022-2}.}
\edit{Complementing this, recent work on the cross-platform migration and diffusion of queer women’s hashtags demonstrates that lesbian communities operate across interconnected platforms, with cultural norms and participation practices adapting as content and users move between spaces~\cite{pan-2015}.}
\edit{These findings suggest that lesbian online spaces function as ecosystems of differentiated sites, rather than as single-purpose platforms. Our findings extend this insight by showing how such differentiation also manifests within a platform, where structurally embedded communities adopt broader, more accessible governance styles, while structurally peripheral communities prioritize protection, specialization, and boundary maintenance.}

Our findings challenges ``one-size-fits-all'' approach to platform governance and moderation tools. Instead, we argue for ecosystem-aware design that recognizes the distinct functional roles of central and peripheral communities.  Central subreddits would benefit from scalable tools for norm-setting, onboarding, and content quality management, such as automated content screens or visual moderation aids, to reduce moderators' burden at scale. By contrast, peripheral subreddits require tools that prioritize risk management, identity protection, and privacy-preserving member control, given their reliance on boundary management.  More broadly, platforms could offer cross-community governance tools. Ecosystem-level tools, such as shared defense mechanisms, cross-community blocklists, or portable reputation signals, could enable collective responses to harassment while reducing redundant labor. Designing for marginalized online communities thus requires shifting from isolated moderation solutions toward infrastructures that support interdependent governance across networks of communities.

\edit{Finally, our findings point to important future directions regarding the integration of AI-enabled features in online community ecosystems. Recent work shows that generative AI can fundamentally reshape participation dynamics, discourse norms, and conflict escalation in online forums, rather than merely supporting individual users~\cite{zeng-2025}. In lesbian and other marginalized communities, such AI-mediated features may amplify existing tensions between openness and protection by altering how content is generated, moderated, and interpreted.}
\edit{From an ecosystem perspective, AI tools are likely to interact differently with structurally central versus peripheral communities: central communities may experience increased visibility and norm-setting power through AI-mediated amplification, while peripheral communities may face risks of misrepresentation, harassment, or boundary erosion. Designers should therefore consider AI not as a neutral add-on, but as a form of community infrastructure whose effects are unevenly distributed across interconnected spaces. This highlights the need for future socially-oriented AI research that attends to ecosystem-level consequences, particularly for marginalized communities.}

\section{Conclusion and Future Work}
We demonstrated that governance is structurally differentiated across the ecosystem, with central subreddits prioritizing visibility through positive definition and peripheral subreddits protecting intimacy through boundary maintenance. Safety encompasses not only content moderation but also participation and identity boundaries. 
This work motivates ecosystem-aware platform design and future research into how users and moderators navigate, sustain, and labor within these differentiated governance roles. Future work will examine how users and moderators navigate this ecosystem, including whether users strategically move between central and peripheral subreddits and how moderators manage the labor and emotional burden of boundary enforcement. We also plan comparative analyses with general-interest subreddit ecosystems to assess whether ecosystem-level governance differentiation is a response to marginalization or a general feature of community scaling.

\section{Appendix}

\begin{figure*}[ht]
    \centering
    \includegraphics[width=\textwidth]{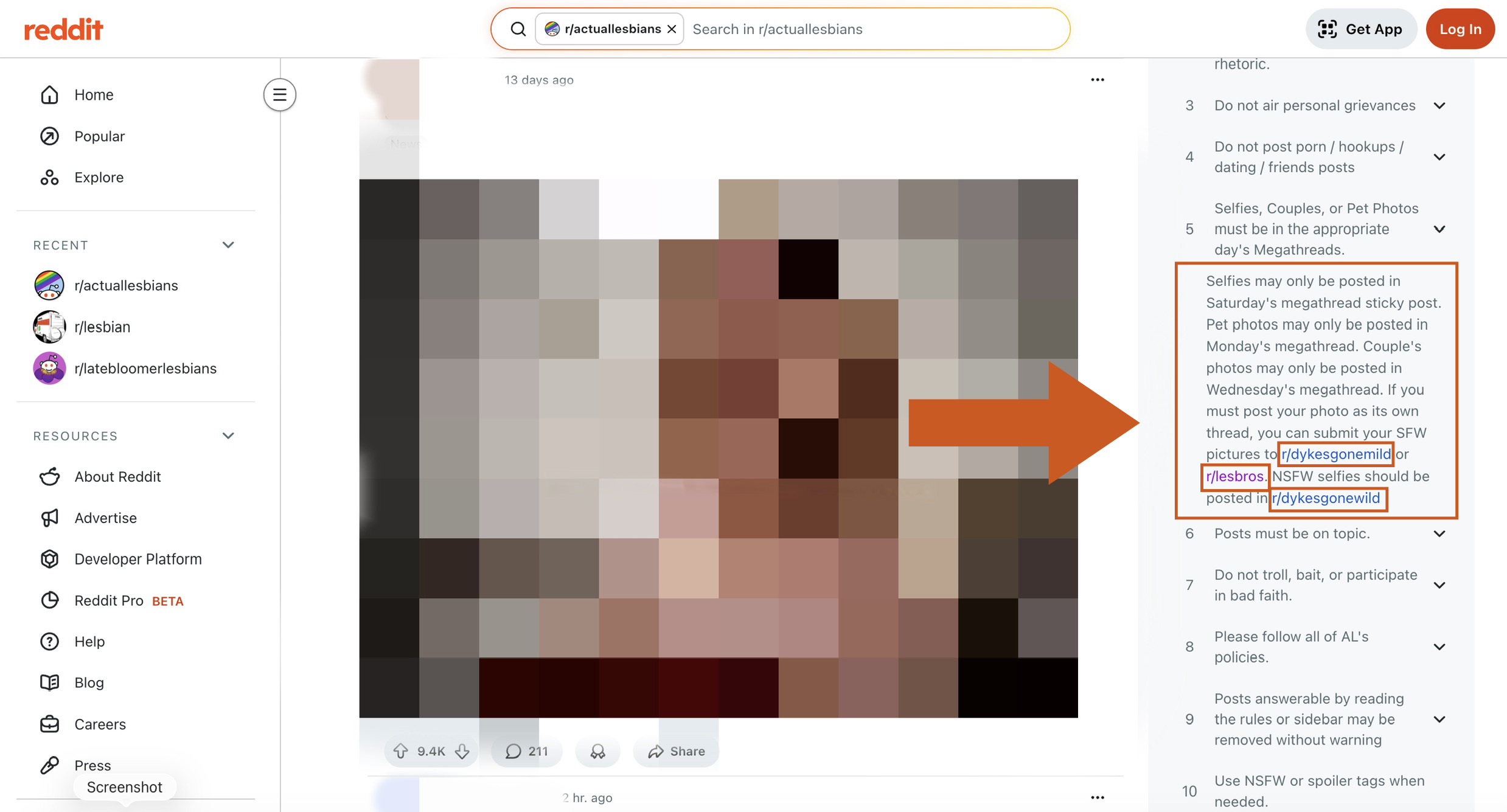}
    \Description{The subreddits mention each other in two ways, one is showing in the screenshot that they cross-link other subreddits in rules.}
    \caption{The lesbian subreddits cross-link to other subreddits in specific rules.}
    \label{fig: rule_mention}
\end{figure*}

\begin{figure*}[ht]
    \centering
    \includegraphics[width=\textwidth]{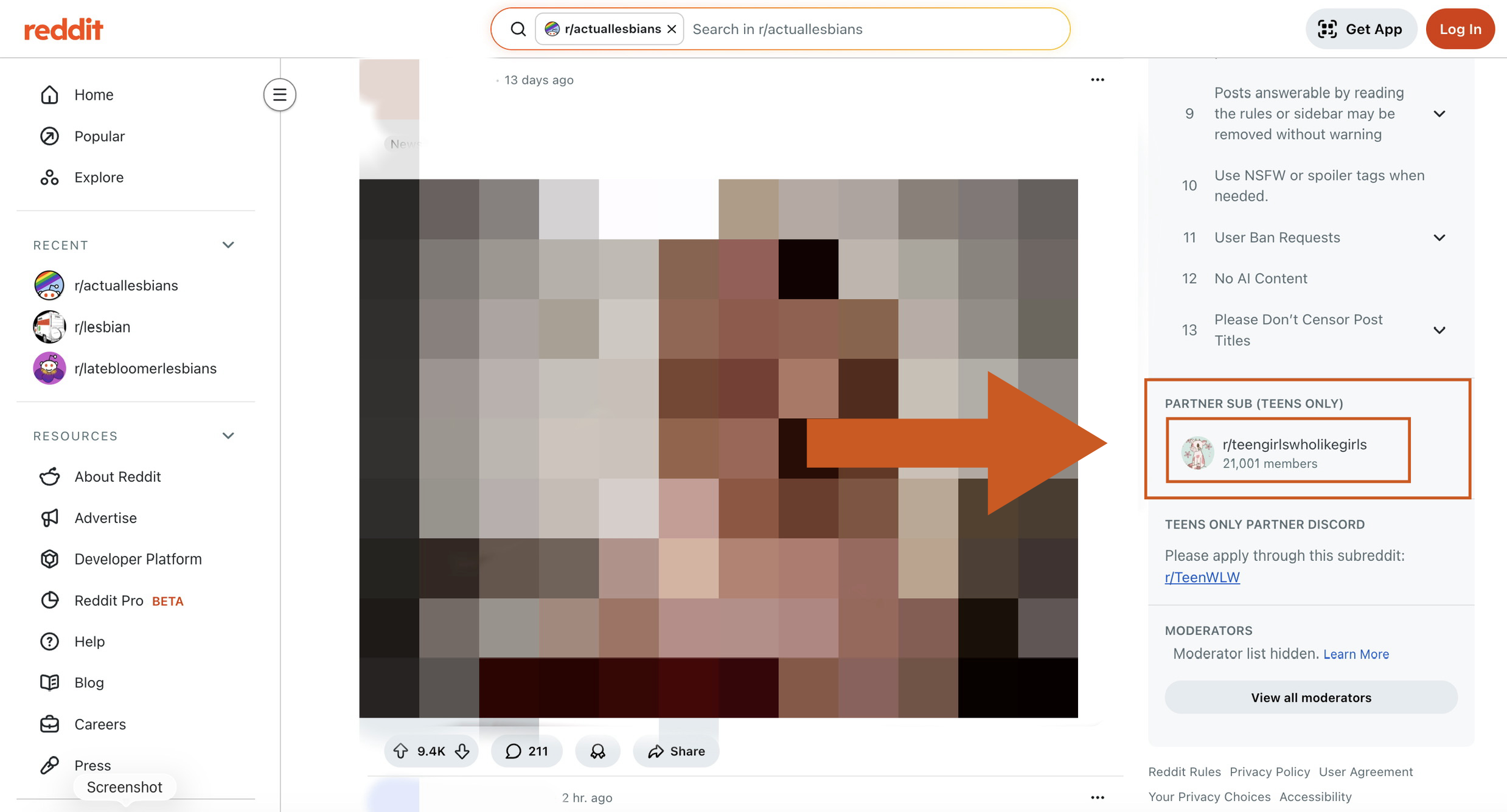}
    \Description{The screenshot shows that subreddits cross-links to other subreddits in ``partner sub'' or ``related subreddits'' widgets on sidebars.}
    \caption{The subreddits cross-link to other subreddits in ``partner sub'' or ``related subreddits'' widgets on sidebars as well.}
    \label{fig:sidebar_mention}
\end{figure*}

These images demonstrate how moderator-curated cross-links appear within the lesbian subreddit ecosystem in two ways: A subreddit explicitly mentions other lesbian subreddits within its governance text. In this example, Rule 5 redirects specific content types (e.g., "SFW pictures" or "NSFW selfies") to designated subreddits. Another way is that a formal "Partner Sub" widget located in the subreddit sidebar which illustrates direct endorsements used to guide users to other lesbian subreddits within the ecosystem.

\bibliographystyle{ACM-Reference-Format}
\bibliography{0References.bib}
\end{document}